\documentclass{aa}
\usepackage{txfonts}

\usepackage{natbib}
\bibpunct[, ]{(}{)}{;}{a}{}{,}
\usepackage{graphicx}
\usepackage[english]{babel}
\usepackage{lscape}

\def\spose#1{\hbox to 0pt{#1\hss}}
\def\lta{\mathrel{\spose{\lower 3pt\hbox{$\mathchar"218$}}
         \raise 2.0pt\hbox{$\mathchar"13C$}}}
\def\gta{\mathrel{\spose{\lower 3pt\hbox{$\mathchar"218$}}
         \raise 2.0pt\hbox{$\mathchar"13E$}}}

\newcommand{\Porb}{\mbox{$P_\mathrm{orb}$}}

\newcommand{\Msun}{\mbox{$\rm M_{\odot}$}}

\newcommand{\Teff}{\mbox{$T_\mathrm{eff}$}}

\newcommand{\Tirr}{\mbox{$T_\mathrm{irr}$}}

\newcommand {\be} {\begin{equation}}
\newcommand {\ee} {\end{equation}}
\newcommand{\beqa}{\begin{eqnarray}}
\newcommand{\eqa}{\end{eqnarray}}
\newcommand{\bea}{\begin{eqnarray}}
\newcommand{\eea}{\end{eqnarray}}
\newcommand {\bc} {\begin{center}}
\newcommand {\ec} {\end{center}}
\newcommand{\nn}{\nonumber}

\begin{document}

\title{Stability of helium accretion discs in ultracompact binaries}
\authorrunning{Lasota et al.}
\titlerunning{Stability of irradiated helium discs}
\author{Jean-Pierre Lasota\inst{1,3}, Guillaume Dubus\inst{1,2}, Katarzyna Kruk\inst{1,3}}

\institute{Institut d'Astrophysique de Paris, UMR 7095 CNRS, UPMC Univ Paris 06,
98bis Boulevard Arago, 75014 Paris, France
\and
Laboratoire d'Astrophysique de Grenoble, UMR 5571 CNRS, Universit\'e J. Fourier, BP 53, 38041 Grenoble, France
\and
Astronomical Observatory, Jagiellonian University, ul. Orla 171, 30-244 Krak\'ow, Poland
\\
\email{lasota@iap.fr} }

\offprints{J.-P. Lasota}

\date{Received / Accepted }

\abstract{Stellar companions of accreting neutron stars in
ultra compact X-ray binaries (UCXBs) are hydrogen-deficient.
Their helium or C/O accretion discs are strongly X-ray
irradiated. Both the chemical composition and irradiation
determine the disc stability with respect to thermal and
viscous perturbations. At shorter periods, UCXBs are
persistent, whereas longer-period systems are mostly
transient.} {To understand this behaviour one has to derive the
stability criteria for X-ray irradiated hydrogen-poor accretion
discs.} {We use a modified and updated version of the Dubus et
al. code describing time-dependent irradiated accretion discs
around compact objects.} {We obtained the relevant stability
criteria and compared the results to observed properties of
UCXBs} {Although the general trend in the stability behaviour
of UCXBs is consistent with the prediction of the disc
instability model, in a few cases the inconsistency of
theoretical predictions with the system observed properties is
weak enough to be attributed to observational and/or
theoretical uncertainties. Two systems might require the
presence of some amount of hydrogen in the donor star.}


\maketitle

\section{Introduction}

It is well known that accretion discs around compact objects
are subject to a thermal--viscous instability at temperatures
corresponding to partial ionization of hydrogen. The model
based on this instability explains the main properties of dwarf
nova and low-mass X-ray binary (LMXB) outbursts \citep[see][for
a review]{lasota01-1}. In the case of transient LMXBs (often
called soft X-ray transients - SXTs) X-ray irradiation of the
accretion disc plays a fundamental role in the stability
criteria and outburst physics
\citep{vp96,kr98,dubetal-99,dubetal-01}. An analogous
instability should be present in ultra-compact binaries
($P_{\rm orb} \lta 60$ min) whose accretion discs are known to
be hydrogen deficient \citep[see e.g.][for recent
reviews]{gijs05,gijs07}. Since some AM CVn binaries (in which
matter is accreted onto a white dwarf) show dwarf-nova type
outbursts and several Ultra Compact X-ray Binaries (UCXBs -
binaries with a neutron-star accreting primary) are transient
\citep[see][]{njs,nj06} it seems that the thermal-viscous
instability is operating also in these systems. Stability
criteria for hydrogen-deficient discs were first obtained by
\citet{smak83} and later by \citet{can84}, \citet{to-97},
\citet{elkhoury} and \citet{menouetal-02}. However, since they
did not include effects of irradiation, they were not
applicable to UCXBs in which irradiation is important
\citep[][but they are of course relevant to AM CVn
stars]{db-03,nj06}.

The formation and evolutionary channels of UCXBs are still
subject to discussion \citep[see e.g.][and reference
therein]{gijs07}. In the so-called white-dwarf channel a double
white-dwarf or neutron-star -- white-dwarf binary evolves due
to angular momentum losses through emission of gravitational
radiation. Such systems evolve from very short to longer
orbital periods with ever decreasing mass-transfer rate
\citep{db-03,detal07,gijs07}. In the second channel a binary
composed of non-degenerate helium star and a white-dwarf or a
neutron-star shrinks first to a minimum period then evolves
back to long periods with a strongly decreasing mass-transfer
rate. Finally, the third scenario involves cataclysmic binaries
or LMXBs with evolved companions. Such systems evolve towards
shorter periods and when mass-loss from the companion uncovers
the helium core their evolution is similar to the previous case
\citep{podsia02}. In all three cases the mass-transfer rate
decreases with orbital period. On the other hand the critical
accretion rate determining the stability of a hot accretion
disc increases with the orbital period. This is because in a
stationary disc, the temperature decreasing with radius and the
critical temperature being practically constant, the stability
is determined by the physical parameters of the outer disc's
ring. The disc radius in turn increases with orbital period.
Therefore one should expect the shorter period UCXBs to be
stable and persistent, while longer period systems should be
transient. This expectation is roughly confirmed by
observations. However, the absence of stability criterion for
irradiated hydrogen-deficient accretion discs has not allowed a
detailed comparison of the evolution models and the Disc
Instability Model (DIM) with observations of UCXBs
\citep{db-03}.

In this article we study the stability properties of irradiated
hydrogen-poor discs and apply our results to UCXBs. In section
\ref{he_stab} we present stability criteria for helium discs and in
the case of no irradiation compare our results with those obtained
by other authors. Application to observed UCXBs is presented in
section \ref{appl} and problems are discussed in the last section
\ref{discussion}. In the Appendix we included updated results for
hydrogen-rich discs.

\section{Thermal-viscous stability of helium discs}
\label{he_stab}

We used an updated and slightly modified version of the code described in
\citet{hameuryetal-98}, \citet{dubetal-99} and \citet{dubetal-01}.

\subsection{Opacities and EOS}

The updated opacities were taken from OPAL
\citep{iglesias_rogers-96} and \citet{alex}. OPAL Rosseland mean
opacities run from $\log T=3.75$ to 8 and the Ferguson et al.
opacities run from $\log T=2.7$ to 4.5. An average of $\log{\kappa}$
is taken in the temperature regime where the tables overlap,
weighted by $\log{T}$, so that the two tables connect smoothly. In
addition to the opacities, the thermodynamical quantities for the
gas mixture (pressure, internal energy, etc.) are calculated
following \citet{bp-69}. Our accretion discs are pure helium
($Y=1$) except  where noted. In the cases where the metal
composition was non-zero ($Z\neq0$), we assume the \citet{grevesse}
metal relative distribution.





\subsection{S-curves}

\begin{figure}
\center
\resizebox{9.0cm}{!}{\includegraphics{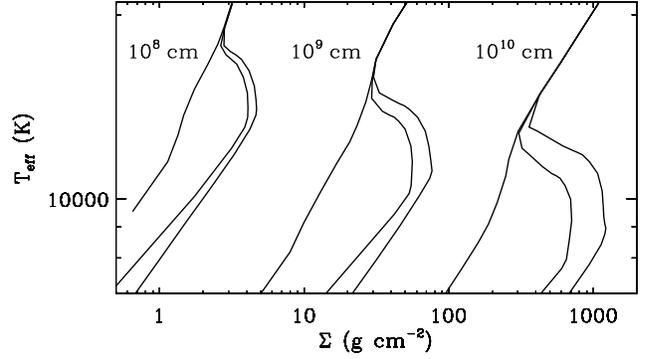}}
\caption{Example S-curves for a pure Helium disk with varying
irradiation temperature $\Tirr$. The plot shows the effective temperature as
a function of the surface density for $\alpha=0.16$. The different
sets of S-curves correspond to different radii $R=10^6$, $10^9$ and $10^{10}$\,cm. For each radius, the irradiation temperature $T_{\rm
  irr}$ is 0\,K, 10 000\,K and 20 000~K. The instability disappears for high irradiation temperatures. }
\label{fig:scurve1}
\end{figure}

The stability condition is found by investigating the local
vertical structure of the disc. The vertical structure  was
calculated following \citet{hameuryetal-98} and
\citet{dubetal-99}. The input parameters are $R_{10}$, $T_{\rm
c}$, $\Tirr$ and $\alpha$ (where $R=R_{10}\,10^{10}\,\rm cm$,
$T_{\rm c}$ the midplane temperature, $\Tirr$ the irradiation
temperature and $\alpha$ the viscosity parameter). The returned
variables are the surface density $\Sigma$ and the effective
temperature $T_{\rm eff}$ for which local thermal equilibrium
is achieved. The solutions form the well-known S-curves in the
($\Sigma$, $T_{\rm eff}$, or $T_{\rm c}$, or $\dot{M}$) plane. We
computed several thousands of such S-curves for $\alpha$ between
$10^{-4}$ and 1, $R$ between $10^6$ and $10^{11}$~cm, $T_{\rm irr}$
between 0 and 25,000 K, $T_{\rm c}$ between 10$^3$ and $10^6$~K.
Example S-curves are shown in Fig. \ref{fig:scurve1}.
The critical values of $\Sigma$ and $T_{\rm eff}$ are given by
the two  inflexion points in the S-curve. Their dependence on the
input parameters can then be approximated by a numerical fit, giving
the critical $\Sigma$, $T_{\rm c}$, $T_{\rm eff}$ as a function of
$\alpha$, $R$ and $T_{\rm irr}$ as in \citet{dubetal-01}.  The
critical mass accretion rate is derived from the above using
\be
\dot{M}=\sigma T_{\rm eff}^4 \frac{8\pi R^3}{3 G M}.
\label{mdot}
\ee
The dependence on $M$ is derived from the dependence on $R$ since
the S-curve is unchanged for a  constant $M/R^3$ ratio.

\subsection{Non-irradiated helium disc: comparison with other calculations and impact of metallicity}
For the non-irradiated case $T_{\rm irr}=0$~K the critical values for a pure helium disc ($Y=1$) are
\beqa
\Sigma_{\rm crit}^{+} & =& 589~\alpha_{0.1}^{-0.78}~R_{10}^{ 1.07}~M_1^{-0.36}\, \rm g\,cm^{-2}\nn\\
T_{\rm c}^{+} & =& 76000~\alpha_{0.1}^{-0.21}~R_{10}^{ 0.08}~M_1^{-0.03}\,\rm K\nn\ \\
T_{\rm eff}^{+} & =& 13100~\alpha_{0.1}^{-0.01}~R_{10}^{-0.08}~M_1^{0.03}\,\rm K \nn\\
\dot{M}_{\rm crit}^{+}& =& 1.05\times 10^{17}~\alpha_{0.1}^{-0.05}~R_{10}^{ 2.69}~M_1^{-0.90}\,\rm g\,s^{-1}\\
\Sigma_{\rm crit}^{-} & =& 1770~\alpha_{0.1}^{-0.83}~R_{10}^{ 1.20}~M_1^{-0.40}\,\rm g\,cm^{-2}\\
T_{\rm c}^{-} & =& 16000~\alpha_{0.1}^{-0.14}~R_{10}^{-0.05}~M_1^{0.02}\,\rm K\nn\\
T_{\rm eff}^{-} & =& 9700~\alpha_{0.1}^{-0.00}~R_{10}^{-0.09}~M_1^{0.03}\,\rm K \nn\\
\dot{M}_{\rm crit}^{-}& =&
3.18\times~10^{16}~\alpha_{0.1}^{-0.01}~R_{10}^{
2.65}~M_1^{-0.88}\,\rm g\,s^{-1}\nn
\label{eq:he_nonirr_crit}
\eqa
where $M_1$ is the mass of the compact object in solar units and
$\alpha=0.1\,\alpha_{0.1}$. The superscript $(+)$ designates a
critical value for hot (``upper branch") discs, whereas $(-)$
corresponds to the cold (``lower branch") discs.

We compare our results with those obtained earlier by other authors.
\citet{menouetal-02} (who use a version of the \citealt{hameuryetal-98} code)
give $\dot{M}_{\rm crit}^{+}$ for a pure-helium disc ($Y=1$)
\be
\dot{M}_{\rm crit}^{+} = 5.9\times
10^{16}~\alpha_{0.1}^{-0.41}~R_{10}^{ 2.62}~M_1^{-0.87}\,\rm
g\,s^{-1}.
\label{eq:menou-he}
\ee
The fit to the critical mass
accretion rate is very close to what we find except for the stronger
dependence on $\alpha$ in the fit of \citet{menouetal-02}. This is
the result of their assumption of a perfect gas equation of state:
as discussed by \citet{menouetal-02}, the dependence on $\alpha$
disappears when convective transport is included, as we have done.

\citet{elkhoury} model the emission from AM CVn stars and present a
few S-curves for some specific values of $\alpha$, $R$ and $M$
assuming  a low hydrogen content. We find good agreement of the
critical $\Teff$ derived from these S-curves (see their Fig. 19).
However, we find our critical $\Sigma$ are higher by $\approx$
50\%.

\citet{smak83} who took $Y=0.98,\,Z=0.02$
obtains $\Sigma_{\rm crit}^{-}\approx 790 \,\rm g\,cm^{-2}$ but in this case
a non-standard way of calculating the cold branch was used
\citep[][and private communication]{smak84}.

As noticed by \citet{to-97}, the critical values depend on
metallicity (see their Fig. 3). This mainly changes the critical
$\Sigma$, especially $\Sigma_{\rm crit}^{-}$. The reason is that at
$\Sigma_{\rm crit}^{-}$ the opacity $\tau\sim 1$ and that changing the
metallicity has a strong effect on the opacities at low temperature.
It is therefore not surprising that our values for $\Sigma_{\rm
crit}^{-}$, found using $Y=1$, differ from  those of
\citet{to-97} whose baseline model uses $Y=0.97\, Z=0.03$. Comparing
the results of our numerical fits (given above) to their $S$-curve
for $Z=0.0004$ (see their Fig. 3), we find that that our values of
$\Sigma_{\rm crit}^{-}$ are within $15\%$ of their values. We also found very
good agreement with their fits  to the critical $\Teff$ and $\Sigma$
(Eqs. 4-7 of their paper) when using the same composition ($Y=0.97\,
Z=0.03$).

One possible consequence of the metallicity dependence of
$\Sigma_{\rm crit}^{-}$ is that running an unstable disc with
$\alpha$ constant can lead to outbursts that have a larger
amplitude when $Z$ is low than when $Z$ is high (because
$\Sigma_{\rm crit}^{-}/\Sigma_{\rm crit}^{+}$ is larger). This
is intriguing as large ratios  of $\Sigma_{\rm
crit}^{-}/\Sigma_{\rm crit}^{+}$ are known to be required to
obtain realistic lightcurves in the framework of the DIM
\citep{smak84}. Typically, this is achieved by lowering
$\alpha$ in quiescence {\em i.e.} assuming less efficient
momentum transport in a cold disc. However, although the ratio
of $\Sigma_{\rm crit}^{-}/\Sigma_{\rm crit}^{+}$ increases with
lower metallicities, the values are still far from the
amplitudes required for realistic lightcurves (about 10-20
compared to a ratio of about 3 here). Therefore this will not
change the conclusion that one needs a lower $\alpha$ in
quiescence for the DIM to work \citep{smak84}.

\subsection{Irradiated helium discs}

In determining the stability of irradiated helium discs we
followed \citet{dubetal-99}: we make a hypothesis on $T_{\rm
irr}$ and find $\dot{M}$ for which the disc of given outer
radius is stable. We assume that $T_{\rm irr}$ is given by
\be
\sigma T_{\rm irr}^4 = {\cal C}\frac{\dot{M}c^2}{4\pi R^2}
\label{eq:tirr}
\ee
where $0\leq \cal C\leq$\ 1 is the irradiation constant defined as in
\citet{dubetal-99}.  The value of $\cal C$ depends upon the
radiative efficiency, irradiation geometry, albedo, irradiation
spectrum etc.

\begin{figure}
\center
\resizebox{9.0cm}{!}{\includegraphics{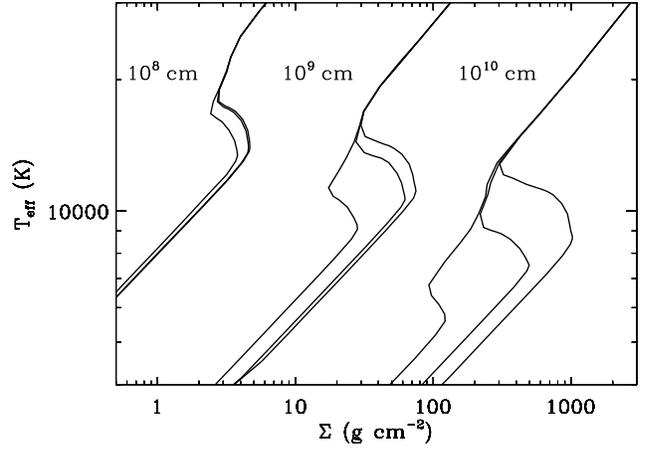}}
\caption{Example S-curves for a pure helium disc with varying
$\cal C$. The plot shows the effective temperature as
a function of the surface density for $\alpha=0.16$. The different
sets of S-curves correspond to different radii $R=10^8$, 10$^9$ and 10$^{10}$\,cm. For each radius, the irradiation
constant $\cal C$ is changed from 10$^{-5}$, to 10$^{-4}$ to 10$^{-3}$. Here, the irradiation temperature varies
along the S-curve. Irradiation has little influence in the inner
regions of the disc but greatly changes the S-curves at the outer radii.}
\label{fig:scurve2}
\end{figure}

With Eq. (\ref{eq:tirr}) we calculate $S$-curves with as input
parameters $R_{10}$, $T_{\rm c}$, ${\cal C}$ and $\alpha$. Here, the
irradiation temperature varies along the S-curve with the mass
accretion rate required for thermal equilibrium. The critical points
are now given as functions of $\alpha$, $R$ and $\cal C$ as in
\citet{dubetal-99}. When one is interested in stability (and
not in time-dependent calculations) $ \cal C$ is assumed to be constant
for a given system. It is therefore more convenient to express the critical
points as a function of $\cal C$ rather than of $\Tirr$  (which itself
depends on $R$ and $\dot{M}$). We computed several thousands of
$S$-curves for $\alpha$ between $10^{-4}$ and 1, $R$ between $10^6$
and $10^{11}$~cm, $T_{\rm c}$ between 10$^3$ and $10^6$~K,
$M_1$ between 0.1 and 100 M$_\odot$ and $\cal C$ between $10^{-10}$
and $10^{-1}$. Examples are given in Fig.~\ref{fig:scurve2}.
\subsection*{Irradiated pure helium disc}

For ${\cal C} \leq 10^{-6}$ one should use the non-irradiated fits given by Eq. (\ref{eq:he_nonirr_crit}).
For ${\cal C}\geq 10^{-6}$ the fits to the critical values are given to a good
approximation (average relative error $\leq$\ 25\% for $\Sigma_{\rm crit}^{+}$,
$\leq$ 10\% for $T_{\rm eff}^{+}$ and $\leq$\ 2\% for $\log \dot{M}_{\rm crit}^{+}$)
\beqa
\Sigma_{\rm crit}^{+}  &=&227
                    ~{\cal C}_{-3}^{-0.16}
                    ~\alpha_{0.1}^{-0.79}
                    ~R_{10}^{ 0.96-0.04\log{\cal C}_{-3}}
                    ~M_1^{-0.25+ 0.04\log{\cal C}_{-3}}\,\rm g\,cm^{-2}\nn \\
T_{\rm c}^{+}         &=& 4480
                   ~{\cal C}_{-3}^{-0.06}
                   ~\alpha_{0.1}^{-0.19}
                   ~R_{10}^{ 0.02-0.01\log{\cal C}_{-3}}
                   ~M_1^{0.02+ 0.01\log{\cal C}_{-3}}\,\rm K\nn\\
T_{\rm eff}^{+}      &=&8730
                   ~{\cal C}_{-3}^{-0.05}
                   ~\alpha_{0.1}^{-0.01}
                   ~R_{10}^{-0.12-0.01\log{\cal C}_{-3}}
                   ~M_1^{ 0.06+ 0.01\log{\cal C}_{-3}}\,\rm K\nn\\
T_{\rm irr}^{+}      &=&22600
                   ~{\cal C}_{-3}^{ 0.20}
                   ~\alpha_{0.1}^{-0.01}
                   ~R_{10}^{ 0.13-0.01\log{\cal C}_{-3}}
                   ~M_1^{-0.19+ 0.01\log{\cal C}_{-3}}\,\rm K\nn\\
\dot{M}_{\rm crit}^{+} &=&2.1~\times 10^{16}
                    ~{\cal C}_{-3}^{-0.22}
                    \alpha_{0.1}^{-0.03-0.01\log{\cal C}_{-3}}
                    ~R_{10}^{ 2.51-0.05\log{\cal C}_{-3}}\nn\\
&&\hskip 3.5cm \cdot M_1^{-0.74+ 0.05\log{\cal C}_{-3}}\,\rm g\,s^{-1}
\label{eq:he_irr_crit}
\eqa
where ${\cal C}= 10^{-3}{\cal C}_{-3}$.

We show only the results for the upper-branch (hot discs).
Since in quiescent SXTs X-ray irradiation is negligible, the
relevant criterion for the lower-branch (cold discs) is the one
without irradiation. For example, according to the DIM in a
quiescent disc the surface density must satisfy $\Sigma <
\Sigma_{\rm crit}^{-}$ \citep[see e.g.][]{lasota01-1}. Since in
such a disc ${\dot M} \sim R^{2.65}$ (Eq.
\ref{eq:he_nonirr_crit}) the ratio of  viscous to irradiating
fluxes $F_{\rm vis}/F_{\rm irr} \sim R^{1.65}$ and
self-irradiation is never important \citep{dubetal-99}.

\subsection*{Irradiated disc with mixed composition}
In order to discuss the influence of the hydrogen depletion we also calculated
the case with X=0.1 and Y=0.9. We find
\beqa
\Sigma_{\rm crit}^{+}  &=&54.0
                    ~{\cal C}_{-3}^{-0.18}
                    ~\alpha_{0.1}^{-0.80}
                    ~R_{10}^{ 0.92-0.04\log{\cal C}_{-3}}
                    ~M_1^{-0.21+ 0.05\log{\cal C}_{-3}}\nn \\
                    && \hskip 6.5cm \rm g\,cm^{-2}\nn \\
T_{\rm c}^{+}         &=&18700
                    ~{\cal C}_{-3}^{-0.07}
                    ~\alpha_{0.1}^{-0.21}
                    ~R_{10}^{-0.05-0.02\log{\cal C}_{-3}}
                    ~M_1^{0.06+ 0.02\log{\cal C}_{-3}}\,\rm K\nn\\
T_{\rm eff}^{+}      &=&4280
                    ~{\cal C}_{-3}^{-0.06}
                    ~\alpha_{0.1}^{-0.02}
                    ~R_{10}^{-0.16-0.02\log{\cal C}_{-3}}
                    ~M_1^{ 0.09+ 0.02\log{\cal C}_{-3}}\,\rm K\nn\\
T_{\rm irr}^{+}      &=&12500
                    ~{\cal C}_{-3}^{ 0.19}
                    ~\alpha_{0.1}^{-0.02}
                    ~R_{10}^{ 0.09-0.02\log{\cal C}_{-3}}
                    ~M_1^{-0.16+ 0.02\log{\cal C}_{-3}}\,\rm K \nn\\
\dot{M}_{\rm crit}^{+} &=&1.9 \times 10^{15}
                    ~{\cal C}_{-3}^{-0.25}
                    ~\alpha_{0.1}^{-0.07+ 0.01\log{\cal C}_{-3}}
                    ~R_{10}^{ 2.38-0.06\log{\cal C}_{-3}}\nn \\
&&\hskip 3.cm \cdot   ~M_1^{-0.64+ 0.07\log{\cal C}_{-3}}
                    \,\rm g\,s^{-1}
\label{eq:mix_irr_crit}
\eqa

\section{Application to ultra-compact binaries}
\label{appl}

The neutron star's stellar companions in Ultra Compact X-ray
Binaries (UCXBs) $P_{\rm orb}\lta 60$ min cannot be
hydrogen-rich stars since such stars would not fit into the
orbit. The 10 confirmed UCXBs (see Table \ref{tab:tab1} have
orbital periods in two ranges: one between $\sim10$  and
$\sim20$ minutes, the second $\sim 40$ and $\sim 50$ minutes.
The four systems below 20 min. are classified as persistent
X-ray sources (except for 4U 1543-624 they are X-ray bursters).
Of the six binaries with periods above 40 min. one is
classified as persistent: 4U 1916-05 which is an X-ray bursting
dipper. The binary 4U 1626-67 (which contains a 7.7 s X-ray
pulsar) used to be classified as persistent but its brightness
is slowly but clearly decaying \citep{krauss07}. The other four
systems are accretion-driven millisecond pulsars and are all
transient systems. In general therefore UCXBs follow the
expected pattern: shorter period are stable, at longer periods
discs become unstable. However, as we shall see, comparing
actual stability criteria with observations makes things a bit
more complicated. In any case one should keep in mind that the
division of UCXBs into persistent and transient might be ``time
dependent". We already mentioned 4U 1626-67 \citep[which should
fade out into quiescence in 2-15 years, see][]{krauss07} but also the
UCXB candidate 1H 1905+000 that was persistent for at least 11
years has now completely disappeared from the X-ray sky
\citep{jonker_1905}.

According to \citet{bp-77} the maximum disc radius can be written as
\be
\frac{R_{\rm D}(\rm max)}{a}=\frac{0.60}{1+q},
\label{eq:rd_bp}
\ee
(valid for $0.03 < q< 1$), where
\be
a = 2.28 \times 10^{9}M_1^{1/3}(1 + q)^{1/3}P_{\rm min}^{2/3}\,\rm cm
\label{eq:separ}
\ee
is the binary separation; $P_{\rm min}$ being the orbital period in minutes.
For convenience we will write the disc outer radius as
\be
R_{\rm D}= 2.28 \times 10^{9}f\,M_1^{1/3}\,P_{\rm min}^{2/3}\,\rm cm
\label{eq:rd_f}
\ee
where
\be
f= \frac{0.60}{(1+q)^{2/3}}.
\label{eq:f}
\ee
\begin{figure}
\center \resizebox{9.0cm}{!}{\includegraphics{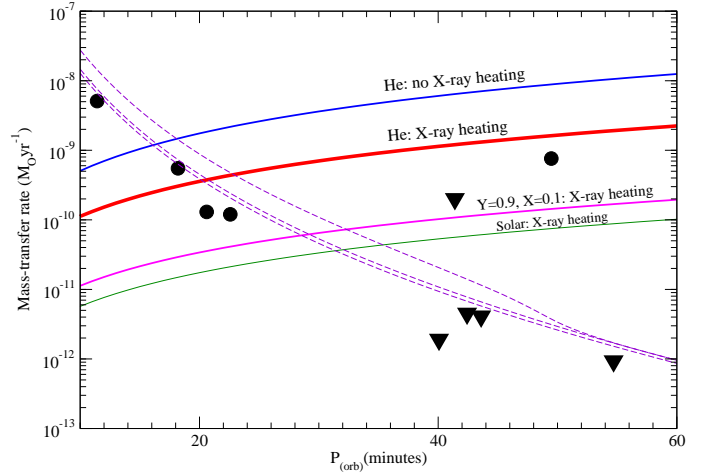}}
\caption{Stability criteria for accretion discs in UCXBs.
\citep[Compare with Fig. 11 in][]{db-03}. Four criteria are plotted
(from top to bottom): (i) for non-irradiated pure-helium discs, (ii)
for irradiated pure-helium discs, (iv) for an irradiated
mixed-composition disc and (v) for an irradiated solar-composition
disc. The accretion rates of persistent UCXBs are marked by circles.
Transient systems are identified by triangles. Dashed lines
correspond to three UCXB evolutionary tracks by Deloye (private
communication). From bottom to top they represent evolution of binaries
with a $1.4\Msun$ neutron star and a secondary with initial mass $M_2$ and
degeneracy parameter $\psi$ (see text) equal respectively: $0.1\Msun$, 3.5;
$0.325\Msun$, 3.0 and $0.325\Msun$, 1.5.}
\label{fig:stability}
\end{figure}
The value of ${\cal C}$ is not known \textsl{a priori}.
Since it is a measure of the fraction of the X-ray luminosity
that heats up the disc it contains information
on the irradiation geometry, X-ray albedo, X-ray
spectrum, etc.  \citet{dubetal-99} found that
the observed optical magnitudes and stability properties
of persistent low mass X-ray binaries were compatible with
a value ${\cal C} \approx 5\times 10^{-4}$. It is not clear that
the same value describes well the properties of UCXBs. Therefore
will take ${\cal C} = 10^{-3}$ as the results do not vary much for
the range $5~10^{-4} \leq {\cal C} \leq 2~10^{-3}$, say.
\begin{table*}
\label{tab:tab1}
\begin{center}
\begin{minipage}{\textwidth}
\caption[UCXBs ($P_{\rm orb}\lta 60$ min)]{UCXBs ($P_{\rm orb}\lta 60$ min)} {\label{tab:ucxb}
}
\setlength{\tabcolsep}{2ex}
\begin{tabular}{lllcl}
\hline\hline\noalign{\smallskip}
System & $P_{\rm orb}(\rm min)$ & $ \dot{M}(\Msun\,\rm y^{-1})$\footnote{see comments in the text}
& Type\footnote{P -- persistent, T -- transient.} &
Comment\footnote{GC -- globular cluster, MSP - millisecond pulsar.}\\
\hline\noalign{\smallskip}
4U 1820-30$^{1}$    & 11.42  & $5.1 \times 10^{-9}$   & P &(in GC)\\
4U 1543-624$^{2}$   & 18.2   & $5.5\times 10^{-10}$  & P &\\
4U 1850-087$^{3}$   & 20.6   & $1.3\times 10^{-10}$  & P &(in GC)\\
M15\, X-2$^{4}$     & 22.58  & $1.2\times 10^{-10}$  & P &(in GC)\\
XTE J1807-294$^{5}$ & 40.07  & $1.9 \times 10^{-12}$  & T &(MSP) \\
4U 1626-67$^{6}$    & 41.4   & $2.0 \times 10^{-10}$   & T & (young pulsar)\\
XTE J1751-305$^{7}$ & 42.42  & $4.5 \times 10^{-12}$  & T &(MSP)\\
XTE J0929-314$^{8}$ & 43.6   & $4.1 \times 10^{-12}$  & T &(MSP)\\
4U 1916-05$^{9}$     & 49.48  & $7.6 \times 10^{-10}$  & P &\\
SWIFT J1756.9-2508$^{10}$ & 54.7 & $9.3 \times 10^{-13}$ & T &(MSP)\\

\noalign{\smallskip}\hline\noalign{\smallskip}
\end{tabular}
\linebreak
\end{minipage}
\end{center}
$^{1}$ \citep{zdz1830},
$^{2}$ \citep{wang_chak},
$^{3}$ \citep{sidetal},
$^{4}$ \citep{dieballm15},
$^{5}$ \citep{markwardt03,falangaetal},
$^{6}$ \citep{krauss07},
$^{7}$ \citep{markwardt02,gierpout},
$^{8}$ \citep{gallowayetal02,jueteal03},
$^{9}$ \citep{juet_chak06},
$^{10}$ \citep{kr07},
\end{table*}
With this value of ${\cal C}$,  from Eqs. (\ref{eq:he_irr_crit}),
(\ref{eq:rd_bp}) and (\ref{eq:separ}), assuming $M_1=1.4$ $\alpha=0.1$, $f=0.6$, one obtains
the following relation between the critical accretion rate and the
orbital period for a pure helium X-ray irradiated disc
\be
\dot{M}_{\rm crit}^{+} =
\begin{array}{cc}
  2.4 \times 10^{-12} f_{0.6}^{2.51} P_{\rm min}^{1.67}\, \Msun\,\rm y^{-1}.
\end{array}
\label{eq:curves_he_irr}
\ee
In the non-irradiated pure-helium disc the equivalent relation reads
\be
\dot{M}_{\rm crit}^{+} =8.2 \times 10^{-12} f_{0.6}^{2.69} P_{\rm min}^{1.79}\, \Msun\,\rm y^{-1}.
\label{eq:curves_he_noirr}
\ee
For an irradiated mixed composition (X=0.1, Y=0.9) the critical accretion rate is
\be
\dot{M}_{\rm crit}^{+} =2.9 \times 10^{-13} f_{0.6}^{2.38} P_{\rm min}^{1.59}\, \Msun\,\rm y^{-1}.
\label{eq:curves_mix_irr}
\ee
and for the solar composition the critical accretion rates for irradiated discs is
\be
\dot{M}_{\rm crit}^{+} =1.4 \times 10^{-13} f_{0.6}^{2.41} P_{\rm min}^{1.61}\, \Msun\,\rm y^{-1}.
\label{eq:curves_h_irr}
\ee
These four criteria are plotted in Fig. \ref{fig:stability}.

\section{Discussion}
\label{discussion}

\citet{db-03} studied the structure and evolution of UCXBs and
showed that for orbital periods $\gta 30$ min the mass-transfer rate
cannot be $\gta 10^{-10} \, \Msun\,\rm y^{-1}$ if they evolved
adiabatically from systems filling their Roche-lobes at
$\Porb\approx 10$ min. As seen in Fig. \ref{fig:stability} one
arrives at a similar conclusion using full stellar models (Deloye,
private communication). These models are parameterized by the total
binary mass and the initial degeneracy parameter of the secondary
given by $\psi=E_{\rm F}/k\,T_c$, where $E_{\rm F}$ is the Fermi
energy and $T_c$ the central temperature \citep[see][]{detal07}. In
general the evolutionary tracks are consistent with the stability
properties of individual systems but there are interesting cases
where this can be questioned.

All UCXBs known to be transient (the millisecond pulsars and
the young pulsar binary 4U 1626-67) have mass transfer rates
well below the stability limit for irradiated helium discs.
However, a word of caution is needed as it is not clear that
the criterion in form of Eq. (\ref{eq:curves_he_irr}) can be
applied to MSP binary systems as their secondary masses are
supposed to be $ < 0.01 \Msun$, As mentioned in
\citet{yungetal06}, for lower values of $q$ matter transferred
from the Roche lobe circularizes on unstable orbits and it is
unclear how disc formation proceeds. The actual outer disc
radius may therefore be smaller (but not by much, see below)
than we have assumed above, which would act to make the systems
stabler than anticipated. On the other hand according to Eq.
(\ref{eq:curves_he_irr}) the maximum outburst luminosity of
helium UCXB MSPs can be expressed as \citep[see
also][]{lasota08-1}
\be
L_{\rm max} \simeq 3.5 \times 10^{37} \left( \frac{P_{\rm
orb}}{1 \rm h}\right)^{1.67}\rm erg\ s^{-1},
\ee
which agrees well with observations. In fact in all UCXBs
mass-ratios could be less than 0.03 but the fact that
luminosities are consistent with the size of a standard disc
model suggests that the stability criterion is applicable even
in such extreme binary systems.

\subsection{Persistent systems}

The mass-transfer rate were calculated by using the
luminosities quoted in the references mentioned in
Table~\ref{tab:tab1} assuming a $1.4\,\Msun$ neutron star with
a 10\,km radius. Among the systems classified as persistent,
two (4U 1820-30 and 4U 1543-624) are stable according to the
stability criterion for irradiated helium discs but three other
binaries (4U 1850-087, M15\, X-2 and 4U 1916-05) should be
transient according to this criterion but apparently are not.
One can try to explain this apparent contradiction in three
ways.

First, the mass-transfer rates (bolometric luminosities) could
be underestimated. Second, the outer disc radius has been
overestimated. Third, the companion are not pure-helium stars
but (still) contain some amount of hydrogen. Let us note that
C/O -- donors would make things even worse for the DIM as the
corresponding stability curves would be {\sl above} the helium
curves \citep{menouetal-02,db-03} because of their higher
ionization potentials.

Considering the first possibility one remarks that of the three
outliers the two with $P_{\rm orb}\sim 20$ min are located in
globular clusters, so one could assume the distance to these
sources is well known. However, the distance to the peculiar
globular cluster NGC 6712 first believed to be 6.8 kpc
\citep{cud88} is now measured to be $\sim 8$ kpc by
\citet{palt6712}. Also in the case of M15 X-2, the
mass-transfer rate based on the X-ray luminosity used in Table
1 could be an underestimate and $\dot{M}\gta 4 \times
10^{-10}$~M$_\odot$\,y$^{-1}$ \citep{dieballm15}.

At first, the second possibility seems to be more promising
since the dependence of stability criterion on the outer disc
radius is very strong. One could think that it is enough to
take a value smaller than the maximum to get rid of the
problem. For example at $P_{\rm orb}\sim 20$\,mn, an outer disc
radius of $0.3\,a$ (a factor $\sim 2$ smaller than assumed
above) is enough for the X-ray heating of a helium disc to
stabilize the system. However, already for $q=0.1$ the
circularization radius is equal to $0.3\,a$ so for the likely
smaller mass-ratio of systems of interest it is not a viable
solution.

It seems finally that the third possibility is the most likely
way out of the stability problem for the two systems. A small
fraction of hydrogen left ($X \gta 0.05$, see Fig.
\ref{fig:stability}) would make them stable. For 4U~1820-30
\citet{cumming03} found that $X\sim 0.1$ is compatible with the
X-ray burst properties of this system. This conclusion,
however, does not apply directly to the two systems of interest
but evolutionary models of \citet{podsia02} would allow them to
have $X\gta 0.1$ (rather paradoxically these models predict
mass-transfer rates higher than observed, in fact in the helium
stability range.)

The evolution of 4U 1916-05 is rather controversial.
\citet{nrj86} suggested that mass-transfer rate deduced for 4U
1916-05 points to an evolved secondaries as donors in this
systems. This would allow the presence of hydrogen. However,
\citet{njs} challenge this conclusion finding a He donor and a
high N abundance. However, it is not clear how this is
consistent with the white-dwarf channel.

Finally, the 20 min systems could have evolved through  the
so-called ``magnetic capture" \citep{sluys05}. In such a (very
unlikely) case they still contain enough hydrogen to fulfill
the stability criterion for $X\lta 0.1$.

\subsection{The unstable UCXBs}

Among the five systems in this category, four are accreting
millisecond pulsars (MSP). The mass-transfer rate was estimated
from the formula
\be
\dot M_{\rm tran}\lta 4\pi \left(\frac{D}{c}\right)^2\,F_{\rm
outb}\frac{t_{\rm outb}}{t_{\rm rec}},
\ee
where $D$ is the distance to the binary, $F_{\rm outb}$ is the
mean bolometric flux during outburst; $t_{\rm outb}$ and
$t_{\rm rec}$ are respectively the outburst duration and the
recurrence time. Except for XTE J1751-305 where the recurrence
time is known to be 2-3 years, $t_{\rm rec}$ was assumed to be
10 years and $\dot M_{\rm tran}$ is therefore an upper limit.
Our estimates our close to those of \citet{wattsetal}. In all
cases the distance to the sources is poorly known so in
addition to the unknown recurrence times this makes
mass-transfer estimates somewhat uncertain. From the point of
view of the instability criterion this is of no importance as
all the MSP are comfortably well below (two orders of
magnitudes) the threshold. Fig. \ref{fig:stability} shows also
evolutionary tracks calculated by Deloye (private
communication) and one might suspect that MSP donors are rather
C/O than He stars \citep[see][]{db-03}.

The case of 4U 1626-67 is different. The neutron star in this
UCXB is a young pulsar and it is transient in a different
sense: its ``outburst" does not last tens of days but tens of
years and it is not clear at all that this behaviour has
anything to do with the thermal-viscous instability of the DIM.
The value of mass-transfer rate given in Table \ref{tab:tab1}
is just a reference value given by Eq. (4) of \citet{krauss07}.
The source has been on since 1977 and seems to be decaying but
the estimate that it will off in 2-15 years is based on dubious
premises. One can say only two things:

The mass-transfer rate of $2.0 \times 10^{-10}
   \,\Msun\,\rm y^{-1}$ is incompatible with evolutionary
   tracks of \citet{db-03}. \citep[See, however,][]{yung08}.

The mass accreted during $\sim 30$ years was $\sim
   5\times 10^{24}$ g (for $D=3$kpc) which is compatible
   with the maximum mass in a cold quiescent helium disc:
\begin{eqnarray}
M_{\rm D, max}\approx  3.5  \times
10^{25}&&\left(\frac{\alpha_{\rm
cold}}{0.01}\right)^{-0.83}\,f_{0.6}^{3.2}\nonumber\\
&&\ \ \times\left( \frac{M_{\rm ns}} {\rm 1.4
M_\odot} \right)^{0.67} \left( \frac{P_{\rm orb}}{1\rm h}
\right)^{2.13}\ {\rm g},
\label{eq:diskmass}
\end{eqnarray}
(see Eq. 3.)
However, the recurrence time could be longer than 1000 years,
say and the mean accretion rate over the cycle $ \lta 2.5\times
10^{-12}\,\Msun\,\rm y^{-1}$, compatible with evolutionary
tracks of \citet{db-03} and \citet{detal07}.

\begin{acknowledgements}
Lars Bilsten inspired this work and contributed invaluable
advice and comments. We thank Chris Deloye for providing UCXB
evolutionary tracks. JPL is grateful to Lev Yungelson for very
helpful discussions and remarks. This work was supported by the
Centre National d'Etudes Spatiales (CNES), the CNRS GDR PCHE
and by the National Science Foundation under Grant No.
PHY05-51164.
\end{acknowledgements}

\begin{appendix}

\newpage

\section{Solar-composition discs}

For comparison with results for helium discs we present here
the fits to critical values for solar composition ($X=0.7$,
$Y=0.28$ and $Z=0.02$)  using the same procedure as described
in the main text.

\subsection*{Non-irradiated solar-composition disc}
\beqa
\Sigma_{\rm crit}^{+} &=&39.9~\alpha_{0.1}^{-0.80}~R_{10}^{1.11}~M_1^{-0.37}\, \rm g\,cm^{-2}\nn\\
T_{\rm c}^{+}         &=&30000~\alpha_{0.1}^{-0.18}~R_{10}^{ 0.04}~M_1^{-0.01}\,\rm K\nn\\
T_{\rm eff}^{+}       &=&6890~R_{10}^{-0.09}~M_1^{0.03}\,\rm K\nn\\
\dot{M}_{\rm crit}^{+}&=&8.07\times10^{15}~\alpha_{0.1}^{-0.01}~R_{10}^{2.64}~M_1^{-0.89}\,\rm g\,s^{-1}\\
\Sigma_{\rm crit}^{-} &=&74.6~\alpha_{0.1}^{-0.83}~R_{10}^{ 1.18}~M_1^{-0.40}\, \rm g\,cm^{-2}\nn\\
T_{\rm c}^{-}         &=&8240~\alpha_{0.1}^{ 0.14}~R_{10}^{-0.10}~M_1^{ 0.04}\,\rm K\nn\\
T_{\rm eff}^{-}       &=&5210~R_{10}^{-0.10}~M_1^{ 0.04}\,\rm K\nn\\
\dot{M}_{\rm crit}^{-}&=&2.64\times10^{15}~\alpha_{0.1}^{0.01}~R_{10}^{ 2.58}~M_1^{-0.85}\,\rm g\,s^{-1}\nn
\label{eq:h_nonirr_crit}
\eqa

\subsection*{Irradiated solar-composition discs}

The fits were obtained with the parameters varied in the range
$10^{-5} \leq \alpha \leq 1$, $10^{-4}\leq R_{10} \leq 10$,
$0.1 \leq M_1 \leq 100$ and $10^{-6} \leq {\cal C} \leq 1$. The
average relative uncertainty on $\log \dot{M}_{\rm crit}^{+}$
is $\leq$ 2\%.
\beqa
\Sigma_{\rm crit}^{+}  &=&8.7
                    ~{\cal C}_{-3}^{-0.28}
                    ~\alpha_{0.1}^{-0.78+ 0.01\log{\cal C}_{-3}}
                    ~R_{10}^{0.92-0.07\log{\cal C}_{-3}}\nn \\
                    &&\hskip 3.5cm \cdot  ~M_1^{-0.19+ 0.06\log{\cal C}_{-3}}\, \rm g\,cm^{-2}\nn \\
T_{\rm c}^{+}          &=&16300
                    ~{\cal C}_{-3}^{-0.10}
                    ~\alpha_{0.1}^{-0.17}
                    ~R_{10}^{-0.03-0.03\log{\cal C}_{-3}}
                    ~M_1^{ 0.05+ 0.02\log{\cal C}_{-3}}\,\rm K \nn \\
T_{\rm eff}^{+} &=&4040
                    ~{\cal C}_{-3}^{-0.09}
                    ~\alpha_{0.1}^{0.01}
                    ~R_{10}^{-0.15-0.02\log{\cal C}_{-3}}
                    ~M_1^{ 0.09+ 0.02\log{\cal C}_{-3}}\,\rm K \nn  \\
T_{\rm irr}^{+}  &=&10500
                    ~{\cal C}_{-3}^{ 0.16}
                    ~\alpha_{0.1}^{0.01}
                    ~R_{10}^{ 0.10-0.02\log{\cal C}_{-3}}
                    ~M_1^{-0.16+ 0.02\log{\cal C}_{-3}}\,\rm K \nn \\
\dot{M}_{\rm crit}^{+} &=&9.5 \times 10^{14}
                    ~{\cal C}_{-3}^{-0.36}
                    ~\alpha_{0.1}^{ 0.04+ 0.01\log{\cal C}_{-3}}
                    ~R_{10}^{2.39-0.10\log{\cal C}_{-3}}\nn\\
                    &&\hskip 3.5cm \cdot  ~M_1^{-0.64+ 0.08\log{\cal C}_{-3}}\,\rm g\,s^{-1}\nn\\
\label{eq_h-irrad} \eqa

\end{appendix}

\end{document}